\documentclass{JINST}
\usepackage{amssymb}
\usepackage{fontenc}
\usepackage{times}
\usepackage{mathptmx}
\usepackage{graphicx}
 \graphicspath{{.}}

\title{High Voltage CMOS Control Interface for 
Astronomy - Grade Charged Coupled Devices}

\author{Elena Martin$^a$\thanks{Corresponding author.},
              Gary Varner$^a$,
		      Aaron Koga$^a$, 
              Larry Ruckman$^a$,
	          Peter Onaka$^b$,
			  John Tonry$^b$,
			  and Aaron Lee$^b$\\

\llap{$^a$} Department of Physics and Astronomy, \\
University of Hawaii,  2505 Correa Rd.,
Honolulu, Hawaii, 96822, USA\\
\llap{$^b$} Institute for Astronomy,\\
 University of Hawaii, Honolulu, 2680 Woodlawn Drive, Hawaii, 96822, USA\\
  E-mail: \email{elena.martin@cnm.es}}

\abstract{The Pan-STARRS telescope consists of an array of 
smaller mirrors viewed by a Giga-pixel arrays of CCDs. 
These focal planes employ 
Orthogonal Transfer CCDs (OTCCDs) to allow on-chip 
image stabilization. Each
OTCCD has advanced logic features 
that are controlled externally. A CMOS 
Interface Device for High Voltage
has been developed to provide the appropiate voltage signal levels 
from a readout and control system designated
STARGRASP. OTCCD chip
output levels range from -3.3V to 
16.7V, with two different output drive strengths required
depending on load capacitance (50pF 
and 1000pF), with 24mA of drive and a rise time on the order of 100ns. Additional 
testing Wilkinson ADC structures have been included in this chip to 
evaluate future functional additions for a next version of the chip.}


\begin{document}

\section{Introduction}

The Panoramic Survey Telescope and Rapid Response System (Pan-STARRS) 
is an innovative design for a wide-field imaging facility being developed 
at the University of Hawaii's Institute for Astronomy.
The combination of small mirrors with very large digital cameras
allows the development and deployment of an economical observing system 
that will be able to survey the entire available sky several times each month.\\

The immediate goal of Pan-STARRS is to discover and characterize Earth-approaching 
objects, both asteroids and comets, that might pose a danger to our planet.
The design of Pan-STARRS is heavily weighted towards its primary purpose, 
which is to detect potentially hazardous objects in the Solar System. 
But the wide-field, repetitive nature of the Pan-STARRS observations 
makes them ideal for a host of other astronomical purposes, ranging 
from Solar System astronomy to cosmology.
There are two features that distinguish Pan-STARRS from other astronomical 
surveys: its ability to map very large areas of sky with great sensitivity 
and its ability to find moving or variable objects. \\

Pan-STARRS will eventually consist of four individual optical systems, each 
with a 1.8 meter diameter mirror observing the same region of sky 
simultaneously. Each mirror will have a 3 degree field of view and 
be equipped with a digital CCD camera containing 1.4 billion pixels. The 
spatial sampling of the sky will be about 0.3 arcseconds. While 
searching for potential killer asteroids in survey mode, Pan-STARRS will cover 
6,000 deg$^2$ per night. The whole available sky as seen from Hawaii 
will be observed 3 times during the dark time in each lunar cycle.
The focal planes will employ \textbf{Orthogonal Transfer CCDs (OTCCDs)} 
that feature on-chip image motion compensation. \\

To evaluate many of the features of the full telescope, 
a prototype system designated
PS1 has been developed. 
PS1 is essentially one quarter of Pan-STARRS. 
It has the same optics and camera design as proposed
for the full version of Pan-STARRS.
PS1 has been built on the site of the south dome of the old LURE 
observatory on Haleakala, Maui.  First light occured in June 2006 
and the telescope was formally dedicated on June 30, 2006. The 
first of the Gigapixel cameras was installed in August 2007.
PS1 serves to test all the technology that is being developed 
for Pan-STARRS, including the telescope design, the cameras, and 
the data reduction software. PS1 will be used to make a full-sky 
survey that will provide astrometric and photometric calibration 
data that will be used for the full Pan-STARRS survey.\\

An overview of the OTA used in the Pan-STARRS
project is found in Section~\ref{cid:background}. Section~\ref{cid:hvc}
introduces the High Voltage control 
chip that has been developed. This chip has two main structures. The first
structure contains the Level Shifting circuitry, described in
Subsection~\ref{slss}, where
test results are presented. The second structure is
a set of Test circuits, which are described in Subsection~\ref{ts}, and where test results for the
ADC are presented.

\section{CID Background} \label{cid:background}
The Orthogonal-Transfer Array (OTA) is a CCD architecture designed to 
provide wide-field tip-tilt correction of astronomical images. This 
device consists of an 8 $\times$ 8 array of small orthogonal-transfer CCDs 
(OTCCD) with independent addressing and readout of each OTCCD. This 
approach enables an optimum tip-tilt correction to be applied 
independently  to each OTCCD across the focal plane. This device was 
developed by MIT  Lincoln Laboratory in support of the Pan-STARRS program (see \cite{HV3}  and \cite{HV4} for more details)
collaborative effort with Semiconductor Technology Associates 
(STA) for the WIYN Observatory.\\

\begin{figure}[htbp]
  \begin{center}
    \includegraphics[width=.4\textwidth, angle=270]{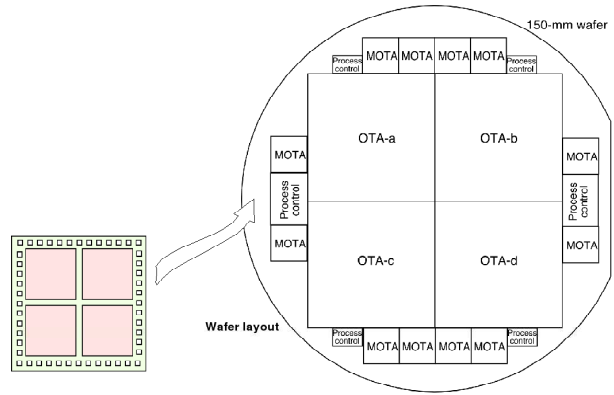}
  \caption{Basic Mini-OTA locations on wafer.
  \label{MiniOTA}}
  \end{center}

\end{figure}

A small, evaluation version of the OTA is called a mini-OTA (MOTA) and consist of a 
2$\times$2 arrays of 
OTCCD cells, shown schematically in Figure~\ref{MiniOTA}. This MOTA was 
included in the wafer layout and has advanced control features of future interest. 
Inclusion of MOTA structures allows the novel nature of the OTCCD devices
to be explored while respecting
the compressed schedule for device deliveries \cite{HV1}, \cite{HV2}. \\

\begin{figure}[htbp]
  \begin{center}
    \includegraphics[width=.4\textwidth, angle=270]{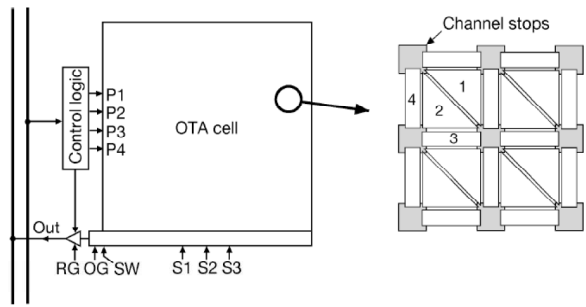}
  \caption{Basic architecture for the OTA cell.\label{OTAarch}}
  \end{center}

\end{figure}

MOTAs provide the opportunity to try advanced features, 
such as two-phase serial registers, 
dual output gates and higher performance designs.
Each cell has an associated control-logic block, as shown 
in Fig.~\ref{OTAarch}. The logic block accepts 3 data bits (D2-D1-D0) 
as inputs and outputs 3 logic levels
 (Z2-Z1-Z0) to control the parallel clock waveforms applied 
to each cell (with Z0 and Z1) and to gate the output of the amplifier onto the video column 
bus OUTn (with signal Z2).The parallel clock voltages applied to each cell are set by four 
active clock levels, P1A-P4A, and to standby voltages, PSH and PSL, all from off-chip circuitry.

\section{CID High Voltage Chip}\label{cid:hvc}

The CMOS Interface Device for High Voltage version 0 (CID0) is a microelectronic chip 
that has been developed using the 0.35$\mu$m High Voltage process from 
AustriaMicroSystems. The main purpose of this chip is to provide the appropiate 
voltage signal levels from the STARGRASP\footnote{http://www.stargrasp.org/} \cite{skymapper} readout and 
control system to the MOTA chip \cite{ota:3,ota:2,ota:1}. 
CID0 also 
has additional signals for output monitoring. Some additional test structures 
have been included in order to evaluate future additions for a next version of the chip. \\

\begin{figure}[htbp]
  \begin{center}
    \includegraphics[width=.4\textwidth, angle=270]{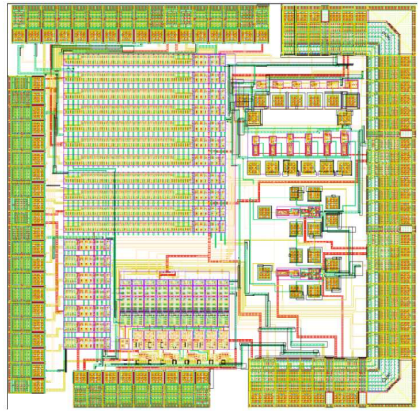}
  \caption{Cadence Layout for the High Voltage Chip.\label{layoutHV}}
  \end{center}

\end{figure}

\begin{figure}[htbp]
  \begin{center}
    \includegraphics[width=.8\textwidth]{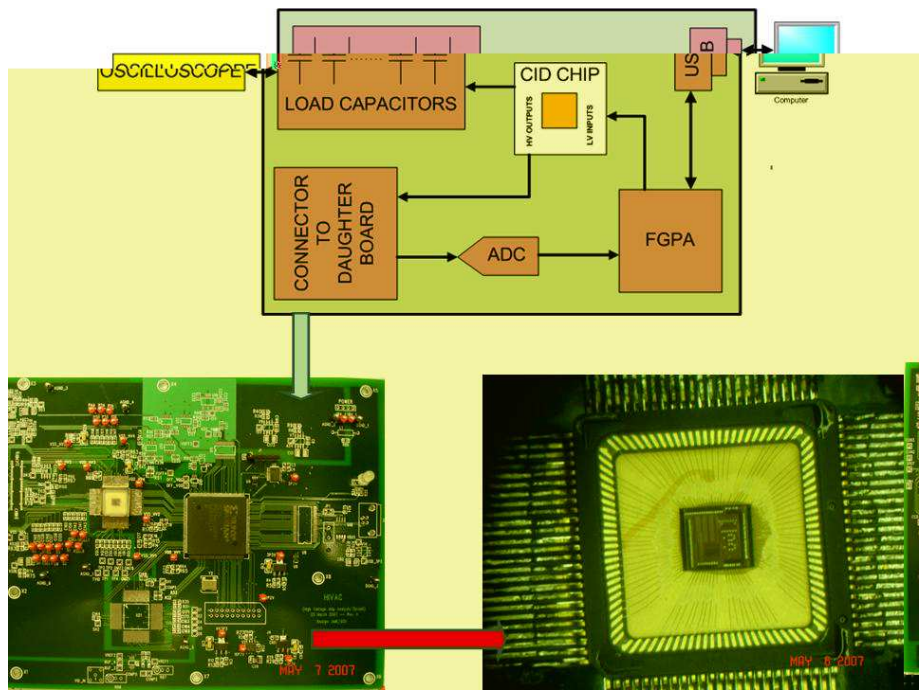}
  \caption{Top image shows the general schematic of the system developed to use the High Voltage Chip, the connector to the daughter board correspond to the CCD connector. The bottom left image shows a photograph of the real board, where the chip can be observed on the left top side of the image. The bottom right image shows a photograph of the chip, where all the wire bondings done (with lengths of $\sim$3mm have been used) can be observed. \label{board_2}}
  \end{center}
\end{figure}

The main structures of this chip are visible in Fig.~\ref{layoutHV}. 
Table~\ref{tabMOTA1} lists the input and 
output signals and Table~\ref{tabMOTA4} the voltage level requirements.  
The external layout dimensions are 3262 x 3218 $\mu m$ (see Fig.~\ref{board_2} to 
see a photograh of the chip at the package). Minimum
pad pitch is 123$\mu m$.\\

\begin{table}[!t]
				\caption[TableMOTA1]{Voltage levels for the MOTA chip}
                 \label{tabMOTA1}
				\centering
                \begin{tabular}{|c|c|c|c|c|c|}
                \hline
                Output& Pin  & V$_{DD}$ & V$_{SS}$ & Ref.& Ref.\\
					       & Name&                 &                  & Level      &Level\\
                \hline \hline
                1          & RSEL1    & V$_{DD\_HV1}$  & V$_{SS\_HV1}$& VDDL           & LREF\\
                2          & RSEL0    & V$_{DD\_HV1}$  & V$_{SS\_HV1}$& VDDL           & LREF\\
                3          & SW          & V$_{DD\_HV2}$  & V$_{SS\_HV2}$& VDDS           & LREF\\
                4          & S3           & V$_{DD\_HV2}$  & V$_{SS\_HV2}$& VDDS           & LREF\\
                5          & S2           & V$_{DD\_HV2}$  & V$_{SS\_HV2}$& VDDS           & LREF\\
                6          & S1           & V$_{DD\_HV2}$  & V$_{SS\_HV2}$& VDDS           & LREF\\
                7          & P4A        & V$_{DD\_HV3}$  & V$_{SS\_HV3}$& VDDH           & LREF\\
                8          & P3A        & V$_{DD\_HV3}$  & V$_{SS\_HV3}$& VDDH           & LREF\\
                9          & P2A        & V$_{DD\_HV3}$  & V$_{SS\_HV3}$& VDDH           & LREF\\
                10        & P1A        & V$_{DD\_HV3}$  & V$_{SS\_HV3}$& VDDH           & LREF\\
                11        & PSH        & V$_{DD\_HV3}$  & V$_{SS\_HV3}$& VDDH           & LREF\\
                12        & PSL        & V$_{DD\_HV3}$  & V$_{SS\_HV3}$& VDDH           & LREF\\
                13        & RG          & V$_{DD\_HV4}$  & V$_{SS\_HV4}$& VDDG           & LREF\\
                14        & CSEL1    & V$_{DD\_HV1}$  & V$_{SS\_HV1}$& VDDL           & LREF\\
                15        & D2          & V$_{DD\_HV1}$  & V$_{SS\_HV1}$& VDDL           & LREF\\
                16        & D1          & V$_{DD\_HV1}$  & V$_{SS\_HV1}$& VDDL           & LREF\\
                17        & D0          & V$_{DD\_HV1}$  & V$_{SS\_HV1}$& VDDL           & LREF\\
                18        & CSEL4    & V$_{DD\_HV1}$  & V$_{SS\_HV1}$& VDDL           & LREF\\
                \hline
                \end{tabular}
\end{table}

\begin{table}[!]
        \begin{center}  
                \caption[TableMOTA4]{Voltage Levels correspondence for Table~\ref{tabMOTA1}.}.
                \label{tabMOTA4}
                \begin{tabular}{|c|c|c|c|c|c|}
                \hline
                $\#$ Pins &         &         & Typ.(V) & Min.(V) & Max.(V)\\
                \hline \hline
                2         & VDD     & 3.3     & 3.3     &         & 4      \\
                2         & VSS     & 0       & 0       & -0.4    &        \\
                2         & VDD\_HV1& LREF+7  & 5       &         & 8      \\
                2         & VSS\_HV1& LREF    & -2      & -10     &        \\
                2         & VDD\_HV2& 10      & 5       &         & 16     \\
                2         & VSS\_HV2& -6      & -3      & -10     &        \\
                2         & VDD\_HV3& LREF+9  & 7       &         & 15     \\
                2         & VSS\_HV3& LREF+1  & -1      & -10     &        \\
                2         & VDD\_HV4& 10      & 5       &         & 16     \\
                2         & VSS\_HV4& -5      & -3      & -10     &        \\
				\hline
                \end{tabular}
        \end{center}
\end{table}

Three main circuit elements included in CID0 occupy different areas 
of the floorplan and use the following areas: the \textbf{Level Shifting 
Subsection} of 1390$\times$1470$\mu$m  for the high 
current drive circuit 
and 432$\times$940$\mu$m for the low current drive circuit, \textbf{Monitor Output 
Signals} area of 945$\times$755$\mu m$ and \textbf{Testing Structures} is
1020$\times$2050$\mu m$.
\subsection{Level Shifting Signal Subsection}\label{slss}
The level shifting circuitry occupies most of the layout.
In all cases, the input signal is a digital signal with input voltage 
levels from 0-3.3V. The output signal  corresponds to the input, but with shifted 
levels and increased drive strength. There are 2 
different capacitive load configurations, 50pF and 
1000pF, and in both cases the circuit has to be able to drive up to 25mA. For either 
case, the rise time is required to be 100ns. \\

A 3-stage topology has been adopted: a pre-amplifier 
that takes the signals from an initial 0-3.3V and translates
to -3.3V to 16.7V, and a 2-stage output buffer, as indicated 
in Fig.~\ref{shifting1}. 
In both cases (high or low load capacitance) the pre-amplifier and the 1$^{st}$ level 
output buffer are equal. The 2$^{nd}$ level buffer determines the final strength of 
the signal. The voltage levels to which the 
various outputs are set are summarized in 
Table~\ref{tabMOTA1} and Table~\ref{tabMOTA4}.\\

\subsubsection{Architecture used for the Level Shift Subsection}

\begin{figure}[htbp]
  \begin{center}
    \includegraphics[width=.4\textwidth, angle=90]{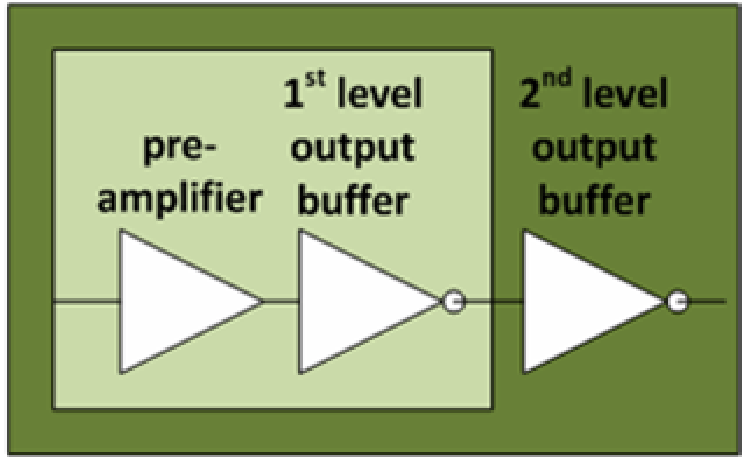}
  \caption{General diagram of the Level Shift 
circuitry.\label{shifting1}}
  \end{center}

\end{figure}

The architecture developed for the Level Shift Subsection consists of 3 different 
parts, as seen in Fig.~\ref{shifting1}.

The pre-amplifier and the 1$^{st}$ level buffer 
convert the signal to the voltage levels needed
at every Voltage Output rail.
The pre-amplifier 
is a basic input-differential comparator 
that converts a low-voltage TTL signal to the extended
 maximum level (-3.3V, 16.7V).
The 1$^{st}$ level buffer shrinks the signal 
as defined in Table~\ref{tabMOTA4}, and uses a group of 3 
inverters that are used as buffers. The size of these transistors are small, using 
$(w/l)_p$=40/5 and $(w/l)_n$=40/2.5.
The third and last stage is a 2$^{nd}$ level buffer, which determines the final strength 
of the output signal drive. It consists of 1 inverter, with 2 different $w/l$ sizes, depending on the 
required output drive strength needed. For loads of 50pF a $(w/l)=200/3$ is used, and 
for loads of 1000pF a $(w/l)=2240/3$ is used. 

\subsubsection{Waveforms obtained}

\begin{figure}[htbp]
  \begin{center}
    \includegraphics[width=.4\textwidth, angle=270]{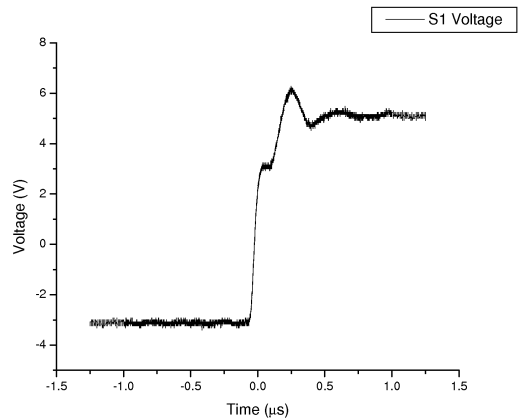}
  \caption{S1 Output voltage signal for a 300pF 
load capacitance.\label{shifting5}}
  \end{center}
\end{figure}

\begin{figure}[htbp]
  \begin{center}
    \includegraphics[width=.4\textwidth, angle=270]{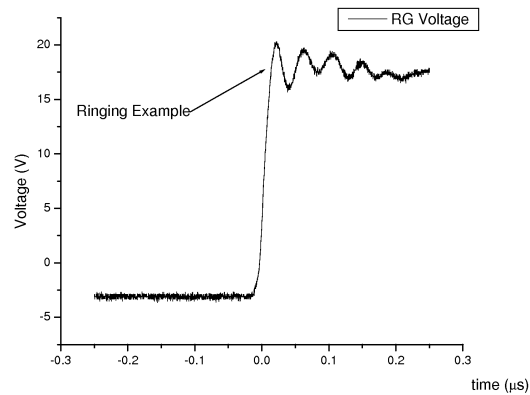}
  \caption{RG Ringing Output Voltage for a 51pF load.\label{shifting7}}
  \end{center}
\end{figure}

The output waveforms observed in the measurements 
show ringing coming from the bonding wire inductance.
Fig.~\ref{shifting5} is a sample Output Voltage for the HV2 rail case. This signal
(S1) is driving a load capacitance of 300pF. 
Fig.~\ref{shifting7}
correspond
to another example of the ringing observed for the RG signal in the case the
output voltage is at HV4 levels and a load capacitor with value of 51pF is
used.

\subsubsection{Rise and Fall Time Measurements}

\begin{figure}[htbp]
  \begin{center}
    \includegraphics[width=.5\textwidth]{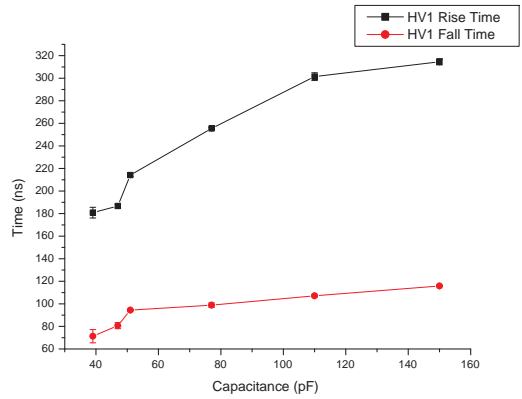}
  \caption{HV1 rise and fall time measurements versus capacitive load.\label{hv1}}
  \end{center}
\end{figure}

\begin{figure}[htbp]
  \begin{center}
    \includegraphics[width=.5\textwidth]{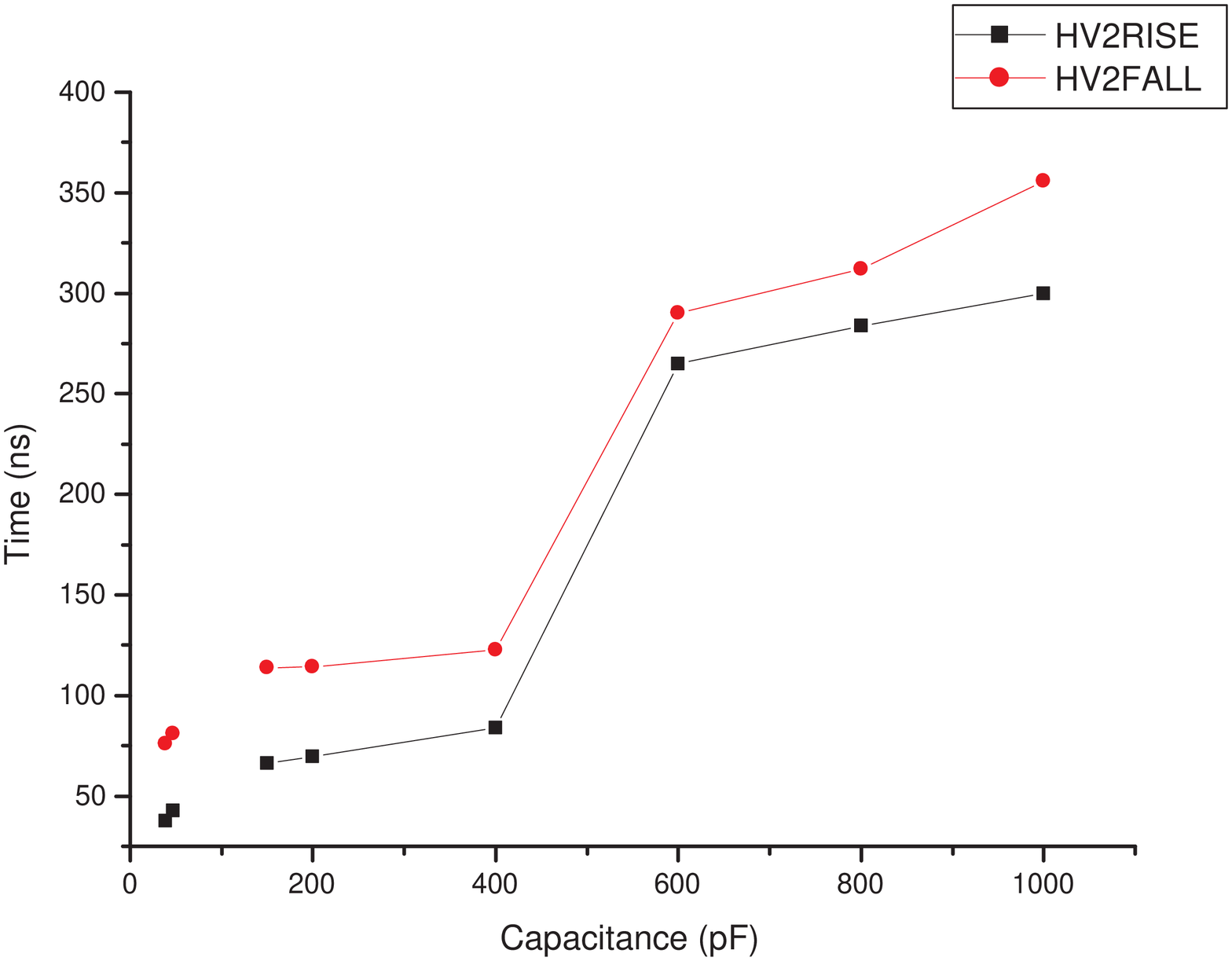}
  \caption{HV2 rise and fall time measurements versus capacitive load.\label{hv2}}
  \end{center}
\end{figure}

\begin{figure}[htbp]
  \begin{center}
    \includegraphics[width=.5\textwidth]{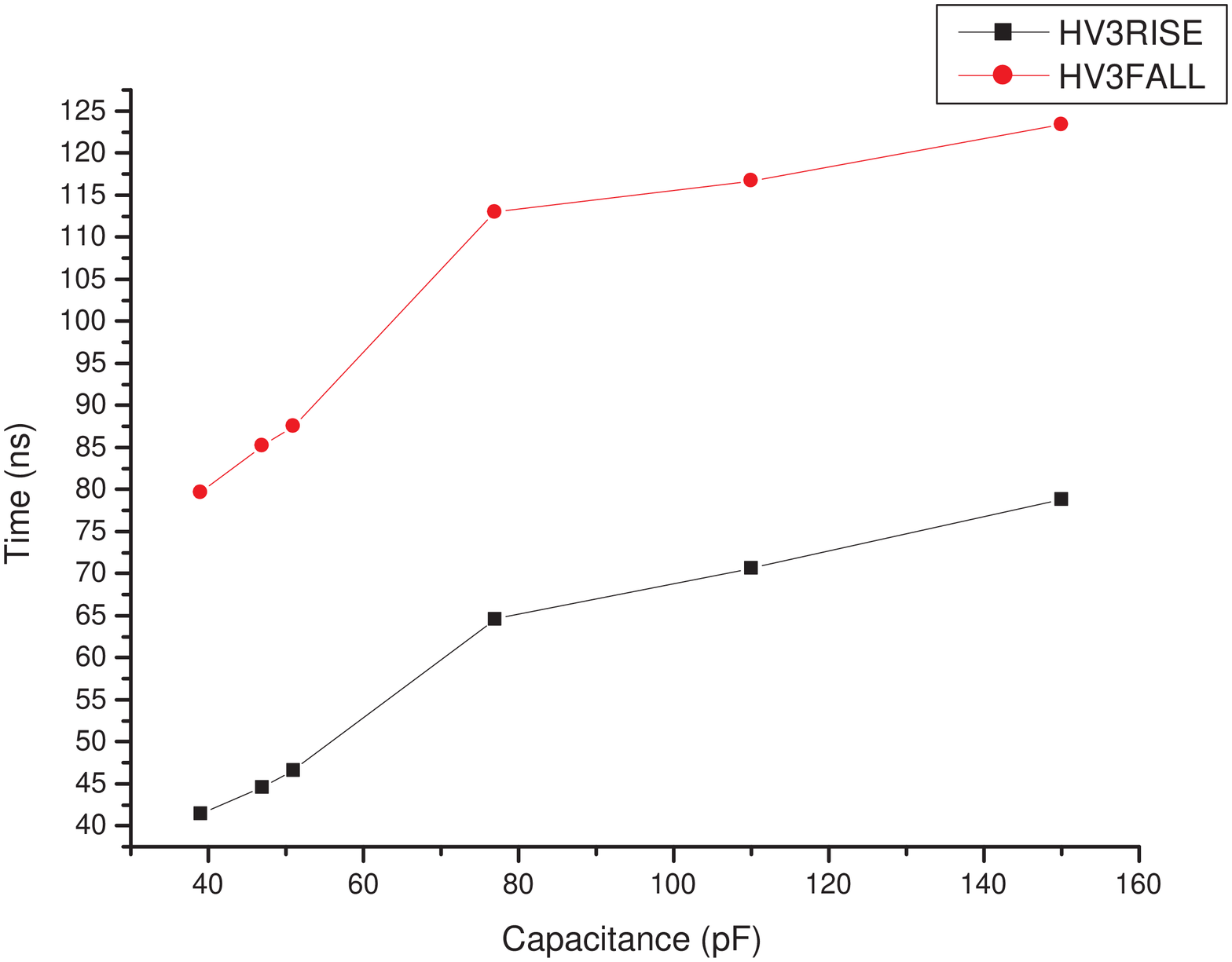}
  \caption{HV3 rise and fall time measurements versus capacitive load.\label{hv3}}
  \end{center}
\end{figure}

\begin{figure}[htbp]
  \begin{center}
    \includegraphics[width=.5\textwidth]{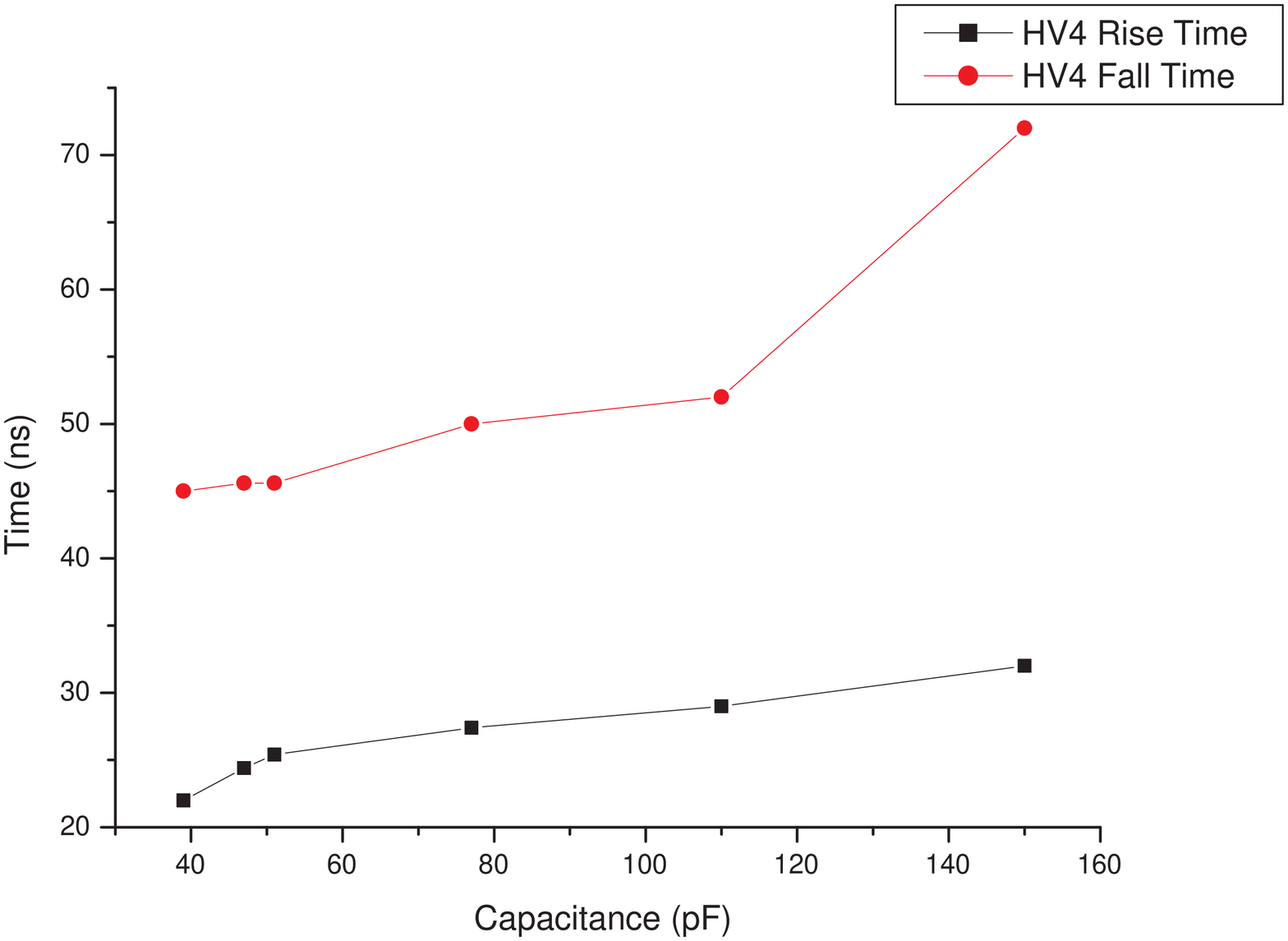}
  \caption{HV4 rise and fall time measurements versus capacitive load.\label{hv4}}
  \end{center}
\end{figure}

Rise and Fall time measurements have been 
performed for voltages HV1-HV4, and these transitions times are shown in 
Fig.~\ref{hv1}-\ref{hv4}, respectively.
These figures show the results of the transition times 
observed for a specific range of load capacitances, 
corresponding to the expected MOTA load 
condition for those signal lines. 

Expected load values for signals of the HV1 are 50pF. 
The rise and fall times are slightly 
larger than expected, with rise times that are $\sim180ns$ for the 
expected output load capacitance. Typical capacitance
load values for output banks HV2, HV3 and HV4 are
1000pF. The measurements are again longer and
higher than had been hoped. The primary reason is that
the maximum drive strength obtained for the pads in this process
corresponds to only 24 mA. This value gives a theoretical 
rise/fall time value for the 1000pF load that is about 3 times the desired value
(100ns).  
The highest values found in Fig.~\ref{hv1} are 300ns for 
the rise time and 350ns for the fall time.
\subsection{Testing Structures}\label{ts}
To evaluate future architectures for inclusion
of a possible ADC in the next CID
version, 3 different testing structures have been included in this chip.

\subsubsection{ADC Ramp circuit}

\begin{figure}[htbp]
  \begin{center}
    \includegraphics[width=.4\textwidth, angle=270]{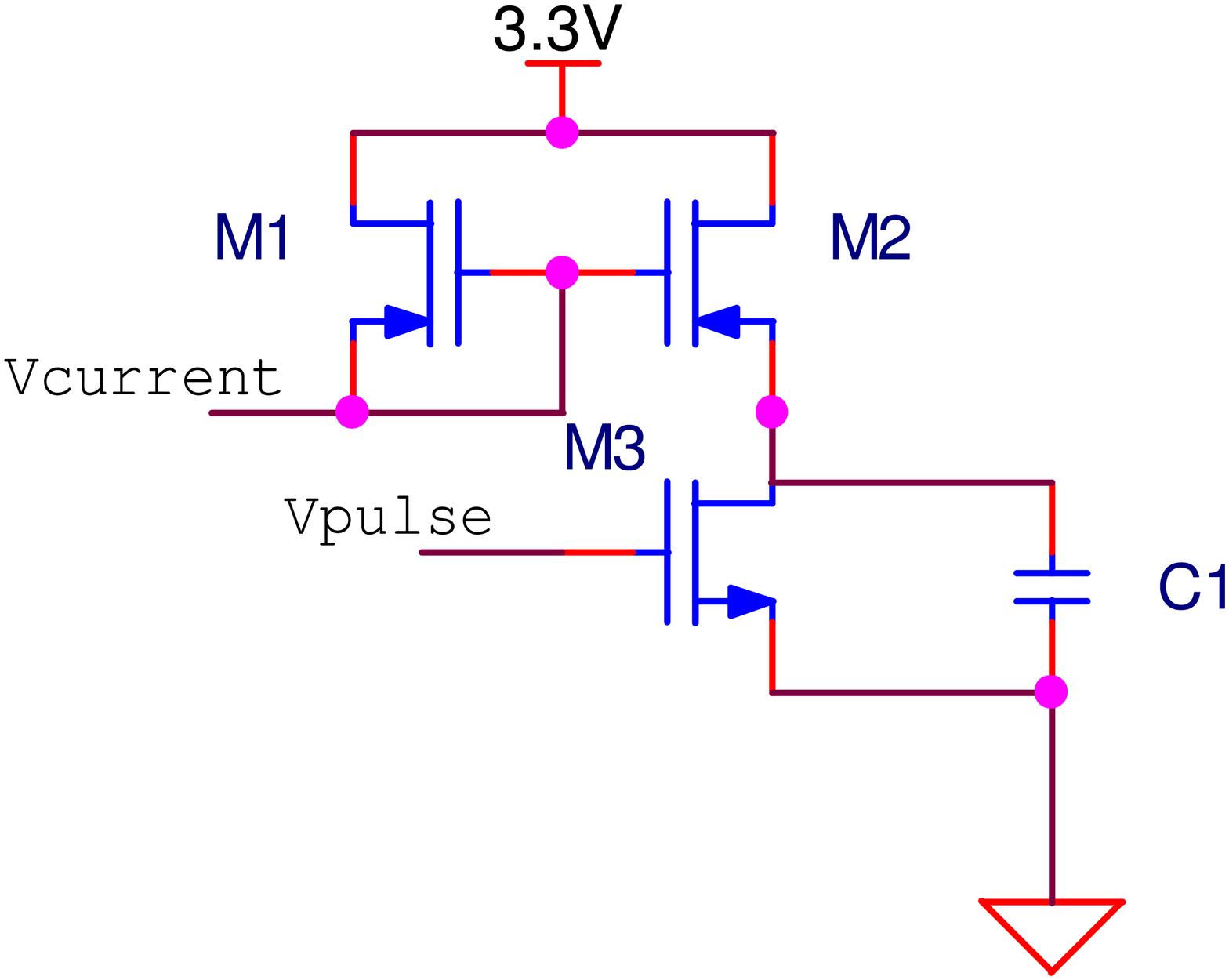}
  \caption{Cadence Schematic for the Ramp Generator.\label{HVRamp1}}
  \end{center}
\end{figure}

 The simple structure shown in Figure~\ref{HVRamp1} 
has been included, where six different values for the 
capacitors have been used (from 0.80pF to 3.12pF).
The capacitor used is C$_{poly}$, formed from poly1 and poly2.

In all configurations the C$_{ramp}$ gets dominated by oscilloscope capacitance (12pF), with $\Delta$V=3.3V and $\Delta$t=2.8ms, showing the current is higher than initially simulated (2.55mA vs 1.0mA)

\subsubsection{ADC Comparator Testing}

\begin{figure}[htbp]
  \begin{center}
    \includegraphics[width=.4\textwidth, angle=270]{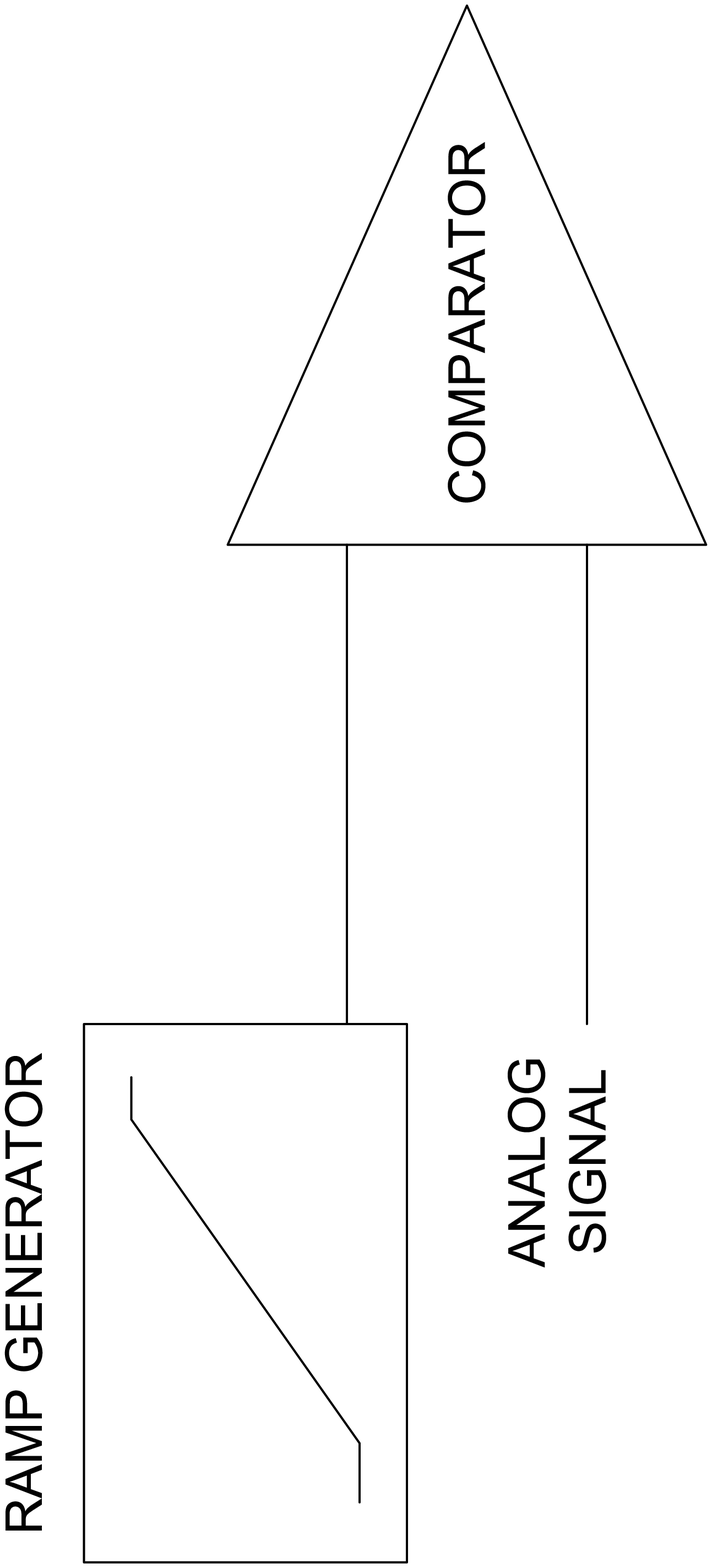}
  \caption{General architecture for the ADC Comparator structure.\label{COMP_TEST}}
  \end{center}
\end{figure}

 This structure includes the previous structure and a comparator.
The comparator is a voltage comparator with a preamplification stage, a decision circuit stage and an output buffer that works at 1MHz. Fig.~\ref{COMP_TEST} shows a 
general schematic that includes the ramp generator that is also present in this 
configuration.

\subsubsection{Sample and Hold Testing}

\begin{figure}[htbp]
  \begin{center}
    \includegraphics[width=.4\textwidth, angle=270]{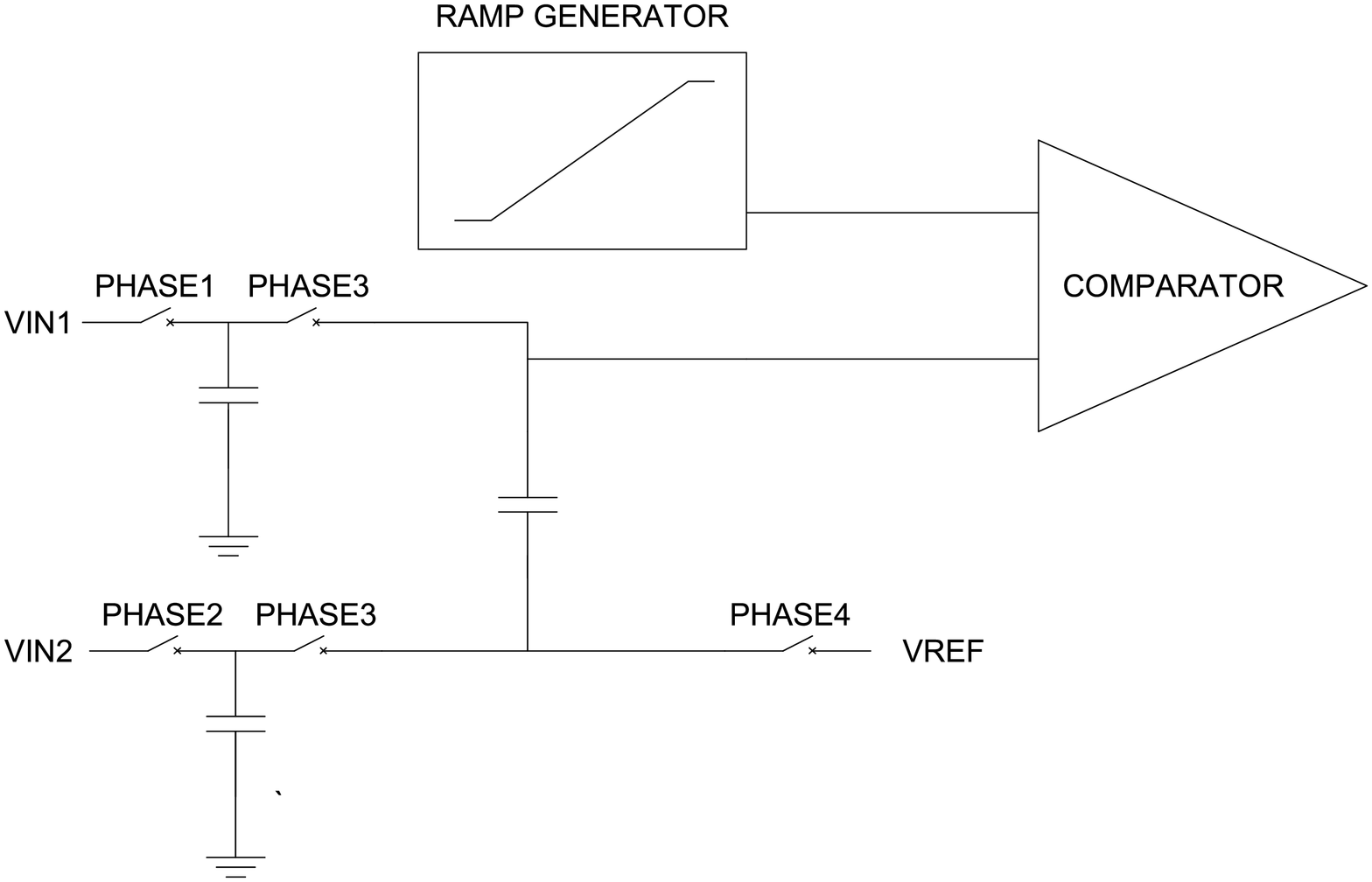}
  \caption{Sample and Hold general architecture.\label{sah}}
  \end{center}
\end{figure}

\begin{figure}[htbp]
  \begin{center}
    \includegraphics[width=.4\textwidth, angle=90]{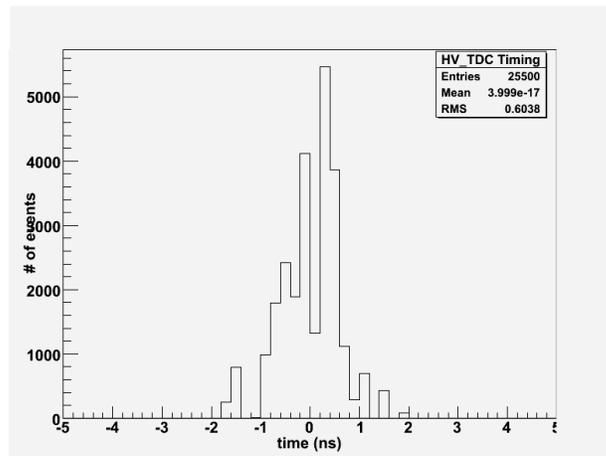}
  \caption{Measurement of the TDC time resolution.\label{HV_TDC}}
  \end{center}
\end{figure}

\begin{figure}[htbp]
  \begin{center}
    \includegraphics[width=.4\textwidth, angle=270]{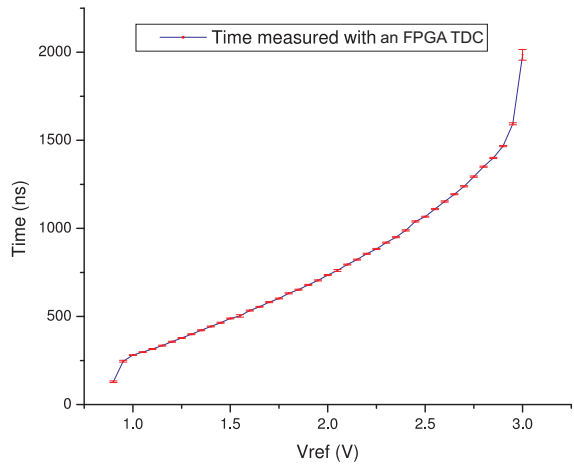}
  \caption{Results for the Wilkinson ADC conversion.\label{graph}}
  \end{center}
\end{figure}

\begin{figure}[htbp]
  \begin{center}
    \includegraphics[width=.4\textwidth, angle=270]{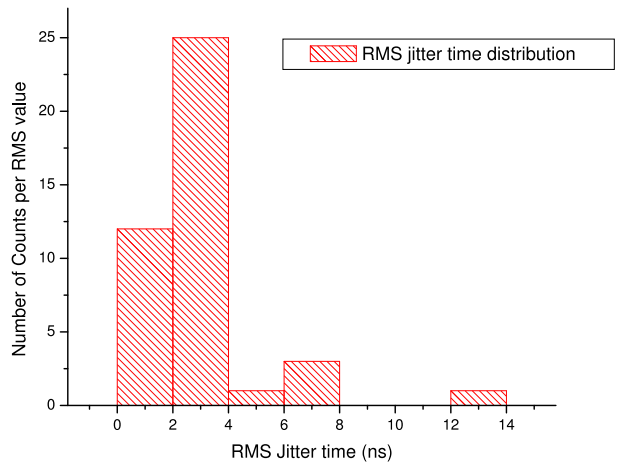}
  \caption{Distribution of RMS noise.\label{rms}}
  \end{center}
\end{figure}

\begin{figure}[htbp]
  \begin{center}
    \includegraphics[width=.5\textwidth]{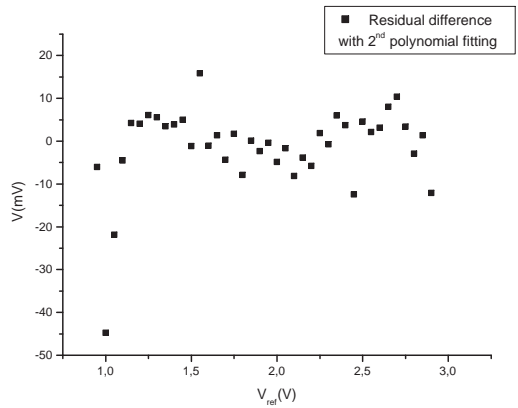}
  \caption{Residual errors in the voltage determination.\label{error}}
  \end{center}
\end{figure}

In this structure
a differential storage of the value is done and then shifted using a reference value. This value is then used for the comparison with the ramp signal, the comparator
output of which time encodes the analog value. Fig.~\ref{sah} ilustrates the general architecture.
To test this structure, an FPGA-based TDC has been used \cite{fpga:tdc},
 with a 2ns time step and a
measured time resolution of 0.604ns, as plotted in
Fig.~\ref{HV_TDC}.
The measured transfer curve is displated in 
Fig.~\ref{graph}. Noise 
coming from the jitter of the FPGA has been substracted from the
total RMS noise, see Fig.~\ref{rms}, according to Eq.~\ref{noise:fpga}.
Fig.~\ref {error} shows the residual errors obtained 
in the voltage determination.

\begin{equation}
\sigma_{measure}^2=\sigma_{edges}^2+\sigma_{FPGA}^2
\label{noise:fpga}
\end{equation}

A fitting of the distribution has been done using a second order polynomial,
with Time = A + B1 $\times$ Vref + B2 $\times$ Vref$^2$, 
with weights given by the error bars. The fitting parameters obtained
are A = 160.75 $\pm$  40.56 ns, B1 = -38.78 
$\pm$ 45.45 ns/V, B2 = 163.83 $\pm$ 11.95 ns/V$^2$. The correlation 
coefficient value is r=0.9955, which promises good performance of the system.

\section{Conclusion}
MOTA CCDs need control signal levels that are much different
than  those typically available for standard logic and need 
to span something like -5V to 20V. To provide this level translation
the CID0 HV CMOS ASIC has 
been fabricated and tested. This chip has been used
to evaluate the 0.35$\mu$m HV process 
from AustriaMicrosystems.
The CID0 chip includes the level shifting circuitry needed for 
MOTA interfacing, as well as 
some additional testing structures. 
The High Voltage Chip is interfaced to the CCD 
using a Flex connector during the test of the prototype.
The final goal of this project is to integrate in the 
same substrate the CCD and the chip, this approach
will also prevent any more ringing problems 
coming from the inductance at the bonding wires used 
in this version.
Observed level shift circuitry performance 
matches expectations for this process, although 
the drive strength of the pads used here is 
not sufficient for a final 10MHz operation. 
In order to meet the design requirements there 
is an option that has been analyzed. This option is to develop 
our own pads, which will show a suitable drive strength value, for 
the time being the CID0 chip will be used with the actual performance.
Wilkinson ADC 
test structures have been evaluated and show promise 
for implementing complex readout structures in close 
proximity and at the high voltage levels typical of 
OTCCD devices.
\acknowledgments

The authors thank the 
Vice Chancellor for Research of 
the University of Hawaii
for his support of this project.

\end{document}